\documentclass[aps,showpacs,prc,nofootinbib,floatfix]{revtex4}

\usepackage{graphicx}

\usepackage{latexsym}
\usepackage{hhline}
\usepackage{amssymb}
\usepackage{wasysym}
\usepackage{epsfig}

\sloppypar

\newcommand{\be}{\begin{equation}}
\newcommand{\ee}{\end{equation}}
\newcommand{\bea}{\begin{eqnarray}}
\newcommand{\eea}{\end{eqnarray}}
\newcommand{\ba}{\begin{array}}
\newcommand{\ea}{\end{array}}
\newcommand{\bi}{\begin{itemize}}
\newcommand{\ei}{\end{itemize}}
\newcommand{\mi}{\mbox i}

\newcommand{\refe}[1]{(\ref{#1})}

\newcommand{\mcl}{{\mathcal L}}

\renewcommand{\slash}{/ \!\!\!\!\,}

\newcommand{\foh}{\frac{1}{2}}

\newcommand{\fth}{\frac{3}{2}}

\begin{document}

\title{On a gauge-invariant interaction of spin-$\fth$ \\
resonances.
\footnote{Supported by DFG}}

\author{V. Shklyar
\footnote{On leave from Far Eastern State University, Sukhanova 8,
                       690600 Vladivostok, Russia}
}
\email{shklyar@theo.physik.uni-giessen.de}
\author{H. Lenske} 
\affiliation{Institut f\"ur Theoretische Physik, Universit\"at Giessen, D-35392
Giessen, Germany}

\begin{abstract}
We show that the gauge-invariant coupling suggested by Pascalutsa removes non-pole 
terms from a spin-$\fth$ propagator only for a specific choice of  free parameter. 
For the general case the problem can be solved by including higher order derivatives
of spin-$\fth$ fields or by modifying  the original coupling. In the latter case 
the obtained Lagrangian  depends on one free parameter pointing to the freedom in choosing 
an 'off-shell' content of the theory. However, the physical observables   must not be 
affected by the  'off-shell' contributions and should not depend on the free parameter 
of the Lagrangian.
\end{abstract}


\maketitle

\section{Introduction}
Since its appearance the theory of higher integer and half-integer spin fields has drawn 
much attention. The physical motivation was to find  an acceptable way of description for 
higher spin baryon and meson resonances. The  interest has mainly been focused on the spin-$\fth$ fields.
This is not surprising because of the role which the $\Delta_{33}(1232)$-isobar plays in $\pi  N$
interaction above the pion production threshold.

In 1941 Rarita and Schwinger (R-S) have suggested a set  of equations  which a field function of 
higher spin should obey \cite{Rarita:1941}.  These constraints should 
directly follow from the corresponding free field Lagrangian of the given spin.
Regardless of the procedure used  the obtained  Lagrangians for free higher-spin fields 
turn out to be always dependent on arbitrary free parameters. In case of spin-$\fth$ fields  
this problem is widely discussed in the literature 
(see e.g.\cite{Benmerrouche:1989uc,Nath:1971wp,Rarita:1941,Singh:1974rc,Fronsdal:1958} ). 
While the free theory is invariant 
under so-called 'point-transformations'  \cite{Benmerrouche:1989uc} inclusion of  
interaction breaks this symmetry. As a result the interaction term  depends on
a parameter of transformation which is commonly known as an 'off-shell' parameter. 
It has been shown that  this dependence cannot be removed
from the full Lagrangian. A critical overview of the problem can be found e.g. 
in \cite{Haberzettl:1998rw,Benmerrouche:1989uc}.

The possibility to construct consistent higher-spin  massless theories has already been 
pointed out by Weinberg and Witten  a long while ago \cite{Weinberg:1980kq}. 
Recently Pascalutsa has shown \cite{Pascalutsa:2000kd,Deser:2000dz,Pascalutsa:1998pw,Pascalutsa:1999zz},
that  by using a gauge invariant coupling for higher spin fields it is possible to remove  
extra-degrees of freedom at interaction vertices.  As a result, the physical observables do not
depend on the off-shell content of the theory. 

It is well known, however, that the  wave equation  for  the free spin-$\fth$ field being written in a general form
depends on one free parameter $A$ (see e.g. \cite{Benmerrouche:1989uc}). The commonly used 
Rarita-Schwinger theory \cite{Rarita:1941} corresponds to the special choice  $A=-1$. While the Pascalutsa-coupling
removes the unwanted degrees of freedom from the  Rarita-Schwinger propagator  it leaves the problem unsolved in 
the more general case $A\neq-1$ resulting  in the appearance of 'off-shell' components, for example  in the  
 $\pi N$ scattering amplitude. 
 Hence, further investigations of the general properties of the interacting spin-$\fth$ fields become 
of great importance.
In this  Letter  we discuss the origin of this  problem and show how it can be solved. We indicate two
alternative approaches. 
The first recipe  consists in constructing a coupling which includes  higher derivatives of the spin-$\fth$
field. Another method  is based on the generalization of the original gauge invariant interaction to arbitrary 
values of $A$. 
In the latter case  the obtained Lagrangian depends on  one free parameter  which also  appears 
in the  free field formalism. However, the physical observables should not depend on this parameter. 
Hence,  the matrix element corresponding to the $\pi N$ scattering at tree level does not contain 
an 'off-shell background'.

\section{\label{FreeF} Spin-$\fth$ field.}
Rarita and Schwinger suggested a set of constraints which the free  spin-$\fth$ field should obey
\cite{Rarita:1941}
\bea
\gamma^{\nu}\psi_{\nu}(x)=0,\nonumber\\ 
\partial^{\nu}\psi_{\nu}(x)=0,
\label{RS}
\eea 
provided that  also the  Dirac equation $(\slash{p}-m)\psi_{\nu}(p)=0$ is fulfilled. 
In the consistent theory  the set of equations eq.\,\refe{RS} should follow from the 
equation of motion obtained from the corresponding Lagrangian. The  Lagrangian of the free spin-$\fth$
field can be written in a general form as follows (see \cite{Benmerrouche:1989uc} and references therein) 

\bea
\mcl^\fth_{0} =\bar \Delta_{\mu}(x)\,\Lambda^{\mu\nu}\,\Delta_{\nu}(x),
\label{lagr0}
\eea 
where $\Delta_{\nu}(x)$ stands for the spin-$\fth$ field and the $\Lambda^{\mu\nu}$-operator is
\bea
\Lambda^{\mu\nu}=&&(\mi \slash \partial -m)g^{\mu\nu}
+\mi A(\gamma^{\mu}\partial^{\nu}+\gamma^{\nu}\partial^{\mu})\nonumber\\
&&+\,\frac{\mi}{2}(3A^2+2A+1)\gamma^{\mu}\slash \partial \gamma^{\nu}\nonumber\\
&&+\,m\,(3A^2+3A+1)\gamma^{\mu}\gamma^{\nu},
\label{L}
\eea 
where $A$ is an arbitrary free parameter subject to the restriction $A\neq-\frac{1}{2}$. The propagator of the 
free spin-$\fth$ field can be obtained  as a solution of the equation, e.g. in momentum space,
\bea
\Lambda_{\mu\rho}(p)\, g^{\rho\sigma} \,G_{\sigma\nu}(p)= g_{\mu\nu}.
\label{prop_eq}
\eea 
The obtained propagator $G_{\sigma\nu}(p)$ can be written as an expansion in terms of the spin projection 
operators  $\mathcal{P}^\fth_{\mu\nu}(p)$, $\mathcal{P}^\foh_{11;\, \mu \nu }$, 
$\mathcal{P}^\foh_{22;\, \mu \nu }$, $\mathcal{P}^\foh_{12;\, \mu \nu }(p)$, and $\mathcal{P}^\foh_{21;\, \mu \nu }(p)$
given in \cite{VanNieuwenhuizen:1981ae}. The first three operators correspond to different irreducible 
representations of spin-vector whereas the  last two ones stand for a mixing between two spin-$\foh$ 
representations, see \cite{VanNieuwenhuizen:1981ae}. In terms of these projectors the propagator is 
represented as
\bea
G_{\mu\nu}(p,A)=\frac{1}{p^2 -m^2+\mi \epsilon}\left( (\slash p + m) \mathcal{P}^\fth_{\mu\nu}(p)
+ \frac{p^2 -m^2}{m\,(2A+1)^2}D^\fth_{\mu\nu}(p)\right),
\label{prop1}
\eea
where the 'off-shell' spin-$\foh$ part  is given by
\bea
D^\fth_{\mu\nu}(p)= - \frac{\left( 1 + 3\,A \right)  
       \left( 2\left (2 + 3\,A\, \right )m +  \left (3\,A+1 \right )\,\slash p \,\right )}{6m}
         \mathcal{P}^\foh_{22;\, \mu \nu }(p) \nonumber\\
 -\frac{\left( 1 + A \right) \,\left( 1 + 3\,A \right)  }{2\, \sqrt{3}\,m} 
                      \, \slash p \left (\mathcal{P}^\foh_{12;\, \mu \nu }(p) -\mathcal{P}^\foh_{21;\, \mu \nu }(p)        
                       \right) \nonumber\\
                 +\frac{\left( 3A^2+3A+1 \right)  }{ \sqrt{3}} 
                    \left (\mathcal{P}^\foh_{12;\, \mu \nu }(p) +\mathcal{P}^\foh_{21;\, \mu \nu }(p)
                      \right) \nonumber\\
-   \frac{ \left( 1 + A \right) 
       \left( 2\,A\,m - \left( 1 + A \right)\, \slash p\, \right)}{2m}
   \mathcal{P}^\foh_{11;\, \mu \nu }(p).
\label{prop}
\eea
While the 'off-shell' part does not contain any poles  it gives rise to the
full propagator in the whole energy-momentum region. The well-known R-S propagator is a special case of 
eqs.\,\refe{prop1},\refe{prop} corresponding  to the  choice $A=-1$, i.e.
\bea
G_{\mu\nu}^{\mathrm{RS}}(p)=\frac{1}{p^2 -m^2+\mi \epsilon}
   \Biggl( (\slash p + m) \mathcal{P}^\fth_{\mu\nu}(p)
       + \frac{p^2 -m^2}{m}
                      \biggl[- \frac{2(\,\slash p +m)}{3m}\mathcal{P}^\foh_{22;\, \mu \nu }(p) \nonumber\\
                                            +\frac{1}{\sqrt{3}}
                                    \left(\mathcal{P}^\foh_{12;\, \mu \nu }(p) +\mathcal{P}^\foh_{21;\, \mu \nu }(p)     
                                    \right )  
                      \biggl ]
   \Biggl).
\label{propRS}
\eea
A commonly used $\Delta N \pi$-coupling is written as 
\mbox{$\mathcal{L}_\mathrm{int}\sim\bar\psi_N \theta(z)^{\nu\mu}\Delta_{\mu} \partial_\nu \pi$}
with \mbox{$\theta^{\mu\nu}(z)=g^{\mu\nu} +z\, \gamma^\mu \gamma^\nu$}. The free  parameter $z$ is used to scale the
'off-shell' contributions to the interaction vertex but does not affect the pole term. In order to remove the 
dependence on the off-shell parameters and 
eliminate the unwanted degrees of freedom Pascalutsa suggested to use a gauge-invariant coupling to the spin-$\fth$ field 
\cite{Pascalutsa:2000kd,Deser:2000dz,Pascalutsa:1998pw,Pascalutsa:1999zz}. The  modified $\Delta N\pi$ interaction 
Lagrangian can be written as follows\footnote{Throughout the paper we omit isospin indices.}
\bea
\mathcal{L}_{P}
=\frac{\mathrm{g}_{\Delta N\pi}}{m_\pi m_N}
\bar \psi_N(x)\gamma_5\gamma_\mu T^{\mu\nu}_\Delta(x)\partial_\nu \pi(x)+\mathrm{h.c.},\nonumber\\
 T^{\mu\nu}_\Delta(x)=\frac{1}{2}\epsilon^{\mu\nu\rho\sigma}
\left(\partial_\rho\Delta_\sigma(x)-\partial_\sigma\Delta_\rho(x)\right),
\label{Pasc}
\eea
where   $\epsilon^{\mu\nu\rho\sigma}$ is the fully antisymmetric Levi-Civita tensor. 
The tensor  $T_{\mu\nu}(x)$ is invariant under the gauge-transformations 
$\Delta_\nu(x)\to \Delta_\nu(x) + \partial_\nu\xi(x)$ where $\xi(x)$ is an arbitrary spinor field. 
Hence, $T_{\mu\nu}(x)$ 
behaves like a 'conserved current' with  the constraint  $\partial_\mu\,  T^{\mu\nu}_\Delta(x)=0$.  The coupling 
defined in eq.\,\refe{Pasc} guarantees that so-called 'off-shell background' 
does not contribute to the physical observables provided that the free spin-$\fth$ propagator is 
chosen in the  special form of eq.\,\refe{propRS} corresponding to $A=-1$. This, however, does not hold in the 
general case for arbitrary 
values of $A$.

The problem can be demonstrated on the example of the $s$- or $u$-channel $\pi N$ scattering amplitude 
calculated from  the corresponding  Feynman graph at
tree level.  The amplitude of interest can be written as
\bea
M=\left(\frac{\mathrm{g}_{\Delta N\pi}}{m_\pi }\right)^2\bar\psi_N(p') \, \left[\,\Gamma_{\mu\rho}(p_\Delta)\, 
\frac{1}{m_N^2}G^{\rho\sigma}_\Delta(p_\Delta)\,\Gamma^\mathrm{\dag}_{\sigma\nu}(p_\Delta)\,\right ]\,\psi_N(p)
\,\,(q_\pi')^\mu\, (q_\pi)^\nu, 
\label{piN}
\eea
 where $p'$$(q_\pi')$ and $p$\,$(q_\pi)$ are momenta of the initial and final nucleon(pion)  correspondingly; $p_\Delta$ stands 
for the momentum of the $\Delta$-isobar and depends on  the channel ($s$- or $u$- ) of interest.  The 'current conserved'
vertex function $\Gamma_{\mu\rho}(p_\Delta)$ reads
\bea
\Gamma_{\mu\rho}(p_\Delta)=\epsilon_{\mu\nu\sigma\rho} \gamma_5 \gamma^\nu p_\Delta^\sigma. 
\label{vertex_Pasc}
\eea
Using the explicit expression  for the free spin-$\fth$ propagator 
the matrix element eq.\,\refe{piN} can be  easily evaluated. For the commonly used  Rarita-Schwinger 
propagator eq.\,\refe{propRS} we have 
\bea
\left[\,\Gamma_{\mu\rho}(p_\Delta)\, G^{\rho\sigma}_\mathrm{RS}(p_\Delta)\,\Gamma^\mathrm{\dag}_{\sigma\nu}(p_\Delta)\,\right ]
=\frac{\slash p + m}{p^2 -m^2}\,\frac{p_\Delta^2}{m_N^2}\, \mathcal{P}^\fth_{\mu\nu}(p_\Delta).
\label{piN_RS}
\eea
This result obtained  first in \cite{Pascalutsa:2000kd,Deser:2000dz,Pascalutsa:1998pw,Pascalutsa:1999zz} 
demonstrates that the interaction chosen in the form of eq.\,\refe{Pasc} removes the lower 
spin components which enter to the original Rarita-Schwinger 
propagator.  A closer inspection shows, that this statement, however, is not true in the general case
as we discuss in the following. 
By choosing the free spin-$\fth$ propagator in the  form  eqs.\,\refe{prop1},\refe{prop} 
with $A\neq-1$,$-\frac{1}{2}$  it is easy to show that the final result contains an extra contribution coming
from  the  $\mathcal{P}^\foh_{11;\, \mu \nu }(p)$ projection operator 
\bea
\left[\,\Gamma_{\mu\rho}(p_\Delta)\, G^{\rho\sigma}(p_\Delta,A)\,\Gamma^\mathrm{\dag}_{\sigma\nu}(p_\Delta)\,\right ]
=\frac{\slash p + m}{p^2 -m^2}\,\frac{p_\Delta^2}{m_N^2}\, \mathcal{P}^\fth_{\mu\nu}(p_\Delta)
+f\left(\mathcal{P}^\foh_{11;\, \mu \nu }(p_\Delta)\right),
\label{piN_Full}
\eea
where $f\left(\mathcal{P}^\foh_{11;\, \mu \nu }(p_\Delta)\right)$ is a function of parameter $A$ and  operator 
$\mathcal{P}^\foh_{11;\, \mu \nu }(p_\Delta)$. The  second (non-pole) term on the right side of eq.\,\refe{piN_Full} represents 
the 'off-shell' spin-$\foh$ background.  By multiplying  e.g. the  right part of eq.\,\refe{piN_Full} by the operator  
$\theta^{\nu\tau}(z)=g^{\nu\tau} +z\, \gamma^\nu \gamma^\tau$  one can see that the obtained expression   will  
again be sensitive to the 'off-shell 'parameter  $z$ -  an unwanted  feature which we want to eliminate 
from the theory. Note, that the  coupling of eq.\,\refe{Pasc} is invented under  assumption that  the interaction 
must support the local symmetries  (gauge invariance) of the free massless R-S formulation, 
which does not hold for $A\neq-1$.

The appearance of the non-pole spin-$\foh$ component  in eq.\,\refe{piN_Full} can be easily traced back. 
The projection operators used in the decomposition of the propagator eq.\,\refe{prop1} correspond to 
various  irreducible representations of the  spin-vector field.  The coupling of eq.\,\refe{Pasc} leads to the 
'current conserved' vertex function  eq.\,\refe{vertex_Pasc}: 
$p_\Delta^\mu \Gamma_{\mu\rho}(p_\Delta)=p_\Delta^\rho \Gamma_{\mu\rho}(p_\Delta)=0$.
Hence, only those terms in eqs.\,\refe{prop1},\refe{prop} give rise  to observables 
which satisfy the condition 
\bea
p^\mu \mathcal{P}^J_{\mathrm{ij};\, \mu \nu }(p)=0,
\label{conservd_current}
\eea
where $J=\fth$,$\foh$ and  $\mathrm{i},\,\mathrm{j}=1,2$.  The  $\mathcal{P}^\fth_{\mu \nu }(p)$-operator fulfills this 
constraint  by definition. The   $\mathcal{P}^\foh_{11;\, \mu \nu }(p)$ projection operator 
corresponds to the $\left[\foh \otimes 1\right]_\foh$ irreducible representation which contains the vector component 
explicitly. Therefore, it  is subject to the condition $p^\mu\,\mathcal{P}^\foh_{11;\, \mu \nu }(p)=0$ too. As a result both 
projection operators contribute to the matrix element eq.\,\refe{piN_Full}.

The problem reported above can be solved in different ways. 
The straightforward one  is to use a coupling with higher order derivatives of the spin-$\fth$ field 
which  explicitly involves the $\mathcal{P}^\fth_{\mu \nu }(p)$ projection operator:
\bea
\frac{\mathrm{g}_{\Delta N\pi}}{m_\pi m_N^4} \bar \psi_N(x) \left[\Box\, 
\mathcal{P}^\fth_{\mu \nu }(\partial) \Delta^\nu(x) \right ]\partial^\mu \pi(x)+\mathrm{h.c.}.
\label{Proj_coupling}
\eea
The use of $\mathcal{P}^\fth_{\mu \nu }(\partial)$ ensures that only  the  spin-$\fth$ part of the 
propagator contributes and the d'Alembert-operator 
guarantees that no other singularities except the mass pole term $(p^2-m^2)^{-1}$ appear in the matrix element.  
Note, that the coupling  written in the form of eq.\,(\ref{Proj_coupling}) restores the  invariance of the full Lagrangian  
under the  'point-like' transformations \mbox{$\Delta_\mu \to \Delta_\mu + z\,\gamma_\mu\gamma^\nu\Delta_\nu$.}

To keep the interaction term in the full Lagrangian as simple as possible we propose here another coupling   
\bea
\mathcal{L}_{I}
=\frac{\mathrm{g}_{\Delta N\pi}}{m_\pi m_N}
\bar \psi_N(x)\, \left [ \,\Gamma_{\nu\eta}\,(A,\partial) 
\,\Delta^\eta(x)\,\right]\,\partial^\nu \pi(x)+\mathrm{h.c.},
\label{Pasc_mod}
\eea
with the modified  vertex operator  $\Gamma_{\nu\eta}\,(A,\partial)$ depending on the parameter $A$:
\bea
\Gamma_{\nu\eta}\,(A,\partial)=\gamma_5\,\gamma^\mu \, \epsilon_{\mu\nu\rho\sigma}\, \theta^{\,\sigma}_{\,\,\eta}(A) \,\partial^\rho,
\label{vert_mod}
\eea 
\bea
\theta_{\sigma\eta}(A)=g_{\sigma\eta}-\frac{A+1}{2}\gamma_\sigma \gamma_\eta.
\label{theta}
\eea
In  momentum space at  $A=-1$   the  vertex function eq.\,\refe{vert_mod} reduces to  that  suggested by Pascalutsa, 
eq.\,\refe{vertex_Pasc}.
The $\theta_{\mu\nu}(A)$-operator has a simple physical meaning:  it relates the Rarita-Schwinger theory to
the general case  of 
arbitrary A. Hence, the R-S propagator, eq.\,\refe{propRS} can be obtained from 
eq.\,\refe{prop1}  by means of the transformation 
$G_{\mu\nu}^{\mathrm{RS}}(p)= \theta_{\mu\rho}(A)\,G_{\rho\sigma}(p,A)\,\theta_{\sigma\nu}(A)$.

Using the coupling eq.\,\refe{Pasc_mod} the final result for the matrix  element of $\pi N$
scattering is independent on the none-pole spin-$\foh$ terms in the full propagator
\bea
\left[\,\Gamma_{\mu\rho}(A,p_\Delta)\, G^{\rho\sigma}(p_\Delta,A)\,\Gamma^\mathrm{\dag}_{\sigma\nu}(A,p_\Delta)\,\right ]
=\frac{\slash p + m}{p^2 -m^2}\,\frac{p_\Delta^2}{m_N^2}\, \mathcal{P}^\fth_{\mu\nu}(p_\Delta)
\label{piN_mod}
\eea
and coincides with that  obtained for the case $A=-1$.
The coupling eq.\,\refe{Pasc_mod} can be written in a more compact form which does not contain the Levi-Civita tensor 
explicitely
\bea
\mathcal{L}_{I}
=\frac{\mi \mathrm{g}_{\Delta N\pi}}{4\,m_\pi m_N}
\,\bar \psi_N(x)\left[ \gamma^{\sigma\rho\nu}\theta_{\sigma\eta}(A)
\partial^\rho \,\Delta^\eta(x)\right]\,\partial^\nu \pi(x)+\mathrm{h.c.},
\eea
where $\gamma^{\sigma\rho\nu}=\{\gamma^{\sigma\rho},\gamma^\nu \}$ and 
$\gamma^{\sigma\rho}=[  \gamma^{\sigma}, \gamma^{\rho}]$ and $\theta_{\sigma\eta}(A)$ is defined in 
eq.\,\refe{theta}.

 Finally, the full Lagrangian for the interacting $\Delta N\pi$ 
fields can be written in the form
\bea
\mcl^{\fth}=\mcl^{\fth}_{0}+\mathcal{L}_{I}+\mathcal{L}_{0}^{\pi}+\mathcal{L}_{0}^{N},
\label{lagr_full}
\eea
where $\mathcal{L}_{0}^{\pi}=\pi(\Box+m^2)\pi$ and $\mathcal{L}_{0}^{N}=\bar \psi_N(i\slash \partial -m)\psi_N$ stand  
for the free Lagrangians of 
pion and nucleon fields, respectively. The free spin-$\fth$ Lagrangian $\mcl^{\fth}_{0}$ and  
$\Delta N\pi$ coupling $\mathcal{L}_{I}$
are given by expressions eq.\,\refe{lagr0} and eq.\,\refe{Pasc_mod}. The Lagrangian eq.\,\refe{lagr_full} depends on one arbitrary 
parameter $A$ which points to the freedom in choosing the 'off-shell' content of the theory. 
Although  $\mcl^{\fth}$ contains one free parameter the physical observables should not depend on it.

A similar conclusion can be made for the electromagnetic coupling. In the notation of \cite{Pascalutsa:2007wz} the 
 generalized coupling can be written as 
\bea
\mathcal{L}_{\gamma N\Delta}=\frac{3\mi e}{2M_N(M_N+M_\Delta)}\bar \psi_N(x) \biggl[ \,
\mathrm{g_M}\,  \theta_{\rho\sigma}(A)\,  \partial_\mu \,\Delta^\sigma \tilde{F}^{\mu\rho} 
+\mi\, \mathrm{g_E}\,\gamma_5\, \theta_{\rho\sigma}(A)\,  \partial_\mu \,\Delta^\sigma F^{\mu\rho} \nonumber\\
-\frac{\mathrm{g_C}}{M_\Delta}\gamma_5\gamma^\rho
   \left (  \theta_{\nu\sigma}(A)\, \partial_\rho \Delta_\sigma  
-  \theta_{\rho\sigma}(A)\, \partial_\nu \Delta_\sigma\right)
\partial_\mu  F^{\mu\nu}\biggl] +\mathrm{h.c.},
\label{lagr_el}
\eea 
where $F^{\mu\nu}$ is the electromagnetic tensor and $\tilde{F}^{\mu\rho}$ is its dual. At $A=-1$  the coupling
eq.\,\refe{lagr_el} reduces to the one  used in  \cite{Pascalutsa:2007wz}.

\section{Summary.}
We have shown that the gauge-invariant $\Delta N\pi$ coupling 
suggested by Pascalutsa for spin-$\fth$  fields removes the 'off-shell' degrees of freedom only for a specific
choice of the spin-$\fth$ propagator but not in the general case. This can be understood by the observation that 
the vertex function  behaves like a 'conserved current'. Hence,  only  those terms in a propagator  are eliminated
which do not have vector components. In the general case the spin-$\fth$ propagator contains a term associated with the
$\left[\foh \otimes 1\right]_\foh$ irreducible representation. The corresponding projection operator 
$\mathcal{P}^\foh_{11;\, \mu \nu }(p)$
does not  contribute to the well-known Rarita-Schwinger propagator  therefore no 'off-shell background' appears.
However, this is not  true in  the general case.  We have shown that the problem can be solved
 by introducing higher order derivatives to the interaction Lagrangian or by  generalizing the 
original $\Delta N\pi$ coupling suggested by Pascalutsa.   In the latter case the full Lagrangian of the 
interacting $\Delta N\pi$ fields depends on one free parameter
which reflects the freedom in choosing an 'off-shell' content of the theory. We have shown 
on the example of $\pi N$ scattering amplitude that  physical observables should not depend on this free parameter.

\begin{acknowledgments}
The work is supported by DFG, contract Le439/7, and SFB/TR16. We are grateful to
Dr. Vladimir Pascalusa and Dr. Jambul Gegelia for  useful remarks on the manuscript.
\end{acknowledgments}

\bibliographystyle{h-physrev3}
\bibliography{tau}

\end{document}